# Band gap control of small bundles of carbon nanotubes using applied electric fields: A density functional theory study

Gunn Kim[1,a], J. Bernholc[2,3], and Young-Kyun Kwon[1]

[1] Department of Physics, Research Institute for Basic Sciences, Kyung Hee University, Seoul 130-701, Republic of Korea

[2] Department of Physics, CHiPS, North Carolina State University, Raleigh, North Carolina 27695-7518, USA

[3] CSMD, Oak Ridge National Laboratory, Oak Ridge, Tennessee 37831-6359, USA

[a]Author to whom correspondence should be addressed. Electronic mail: gunnkim@khu.ac.kr.

**Electrostatic screening between carbon nanotubes (CNTs) in a small CNT bundle leads to a switching behavior induced by electric field perpendicular to the bundle axis. Using a first-principles method, we investigate the electronic structures of bundles consisting of two or three CNTs and the effects of the electric field applied perpendicular to the bundle axis. The applied field causes band gap closure in semiconducting bundles, while a gap opening occurs in metallic ones, which enables considerable modulation of bundle conductivity. The modulation effect originates from symmetry breaking due to electrostatic screening between the adjacent tube walls.**

Due to van der Waals interactions between neighboring nanotubes, carbon nanotube (CNT) bundles are readily produced, and small bundles comprising a few tubes are often observed in chemical vapor deposition growth. Investigations of electrical transport in tube bundles have uncovered many interesting features such as single electron transport and metallic resistivity. For CNT-based electronics, it is necessary to account for the role of cooperative effects of nanotube bundles, and of their effect on transport properties in response to external electric fields. For example, in a recent investigation of a field effect transistor based on suspended CNTs, the *I-V* characteristics of a small CNT bundle were quite different from those of a combination of isolated single-walled CNTs, e.g., in the bundle the off-state current was not completely independent of the gate bias.[1]

In order to understand the response of small bundles of CNTs to applied gate bias or an external electric field, we perform first-principles calculations for bundles consisting of two or three tubes. We find that the electronic properties of the small CNT bundles can be explained by variation in the band gap of the bundle due to an applied electric field. These effects are different from those occurring in isolated single-walled CNTs, demonstrating that electronic transport in a small bundle can be dramatically influenced by cooperative effects and the direction of the applied external field. The cooperative effects are associated with electrostatic screening in the region between the adjacent CNTs.

The *ab initio* electronic structure calculations for the bundles are carried out using density functional theory (DFT) within the local density approximation[2] for the exchange-correlation energy. Some calculations are repeated using the generalized gradient approximation and lead to essentially the same results. Norm-conserving Troullier–Martins pseudopotentials[3] are employed and the wave functions are expanded using a numerical atomic orbital basis set in the SIESTA code[4,5] with an energy cutoff for the real space mesh of 250 Ry. The wave functions are expanded with a double-$\zeta$ plus polarization basis set.[4,5] A periodic saw-toothlike potential is used to create a homogeneous electric field across the tube region. Since chiral tubes would have an unacceptably large repeat period, we choose a semiconducting zigzag (11,0) CNT with a band gap of 0.8 eV. CNTs with different chiralities but comparable diameters are expected to show similar electronic features. The intertube distance between nanotubes in the bundle changed from the initial value of 3.34 Å (the interlayer distance of graphite) to around 3.0 Å. This work is also repeated using the OPENMX code.[6,7,8] General features of the computational results from the two packages are almost identical. In the cases of three tube bundles, all three CNTs interact with one another through screening charges. They also show dramatic energy band gap modification due to the applied electric field (not shown here).

We investigate the effect of an external electric field on the electronic property of two semiconducting tube bundles. It was shown in a previous study that a bundle consisting of two metallic single-walled carbon nanotubes (SWNTs) opens a gap because of intertube interaction.[9] Figures 1a,1b show the band structures of a (11,0)–(11,0) SWNT bundle in an external electric field of 0.2 V/Å perpendicular to the tube axis. The band structures clearly reveal the dramatic and direction-sensitive change in the band gap upon application of the field. While the band gap is about 0.6 eV when each field line passes through only one nanotube [Fig. 1a], it is reduced to $\sim$ eV 0.05 when field lines pass through both tubes [Fig. 1b]. Clearly, the electric field enhances the intertube coupling by polarizing the charge density. As shown in Fig. 1a, when field lines do not penetrate both tubes, the electronic structure is not significantly changed for a field of 0.2 V/Å, when compared to the zero field case, where the band gap of the bundle is almost the same (0.6 versus 0.7 eV).

Since the band gap scales as the square inverse of the diameter of the tube and the perturbation due to a transverse electric field is proportional to the diameter,[10] one expects the band gap of a bundle of semiconducting SWNTs to disappear when a sufficiently strong transverse field is applied. Our calculations reveal that the critical transverse field strength in the *x* direction is between 0.3 and 0.4 V/Å for the semiconductor-metal transition of the (11,0)–(11,0) CNT bundle. For $E_x$ = 0.4 V/Å, the (11,0)–(11,0) CNT bundle becomes metallic as clearly seen in Fig. 1c. Figure 2 shows the density of states for various electric fields and we can see the same features of energy band gap changes due to the intensity and the orientation of the applied electric field as in Fig. 1.

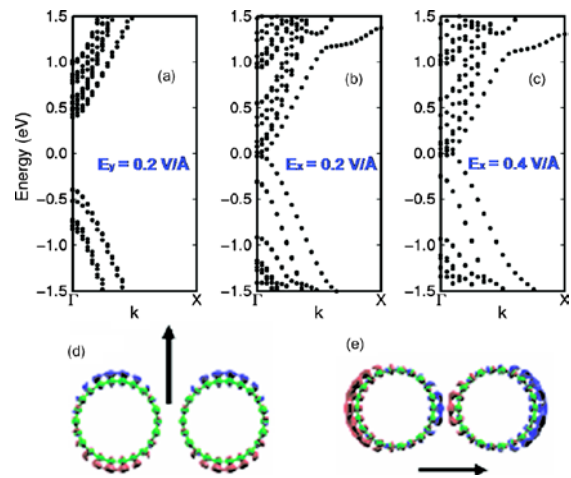

**Figure 1**

Band structures of (11,0)–(11,0) two-tube bundles under applied electric field and their charge density difference plots. As shown in (b) and (c), due to a cooperative effect enhanced by the field, the band gap is reduced when $E_x$ is applied. The arrows in (d) and (e) represent the directions of external electric fields.

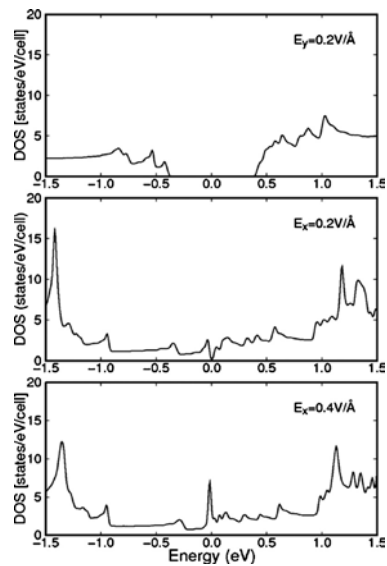

**Figure 2**

Density of states of the (11,0)–(11,0) CNT bundle under applied electric field. As the intensity of $E_x$ becomes larger, the energy band gap decreases. For $E_x$ = 0.4 V/Å, the CNT bundle clearly show metallic behavior.

For an individual (11,0) CNT, much larger field (>V/Å) is required for an appreciable band gap change. In this work, we concentrate on the phenomenon of a drastic band gap change in the carbon nanotubes under the external electric fields, and do not obtain more accurate values of band gaps using the GW correction. We believe that the GW correction or other methods for obtaining more accurate results will also show these interesting band gap changes in CNTs of a small CNT bundle under external electric field. In Figs. 1d,1e, the isodensity surface plots of charge density differences for the two cases obviously show that the inner surfaces of the two-tube bundle exhibit a much smaller charge displacement than the outer surfaces when each field line passes through both tubes. The screening effect of the field in the intertube region is responsible for the large change in the band structure.

Figures 3a,3b show two-dimensional maps of the electric potentials of the (11,0)–(11,0) CNT bundle under an applied electric field. In the regions where the spacing of equipotential lines is wide, i.e., in the region between the two tube walls, the electric field is weak due to screening effects. Since the potential seen by the electrons is severely deformed, the band structure of the SWNT bundle changes. Owing to the energy level splitting of adjacent nanotubes, the band gap of the (11,0)–(11,0) tube bundle is smaller than that of a single (11,0) CNT.

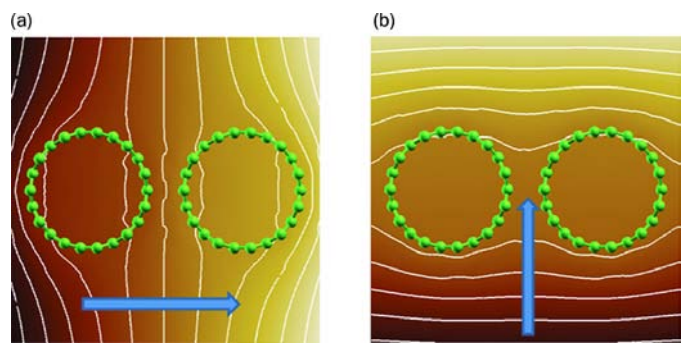

**Figure 3**

Two-dimensional map of the electric potential in the (11,0)–(11,0) CNT bundle under applied electric field. In the region between the two tube walls, the electric field is weaker due to screening by the walls.

In general, a stronger intertube coupling would result in a stronger perturbation of the band structure in a bundle.[11,12,13] Therefore, the electronic structure of a CNT bundle could be dramatically affected by the combined effects of external fields and interactions with the substrate and/or adjacent CNTs. This is especially true for bundles of large-diameter SWNTs, where the van der Waals attraction between neighboring SWNTs will result in substantial radial deformations. A transmission electron microscopy measurement shows the deformation of 9% for the maximum radial contraction in the case of the (19,11) CNT with the diameter of nm ~ 2.[14] In a previous work, we found that radial deformation may give rise to a noticeable change in the band gap of the CNT bundle.[15] In the present study, it is also found that the two (11,0) CNTs in the bundle undergo small radial deformation due to electrostatic interaction between two CNTs caused by the induced polarization under the applied field. This radial deformation slightly breaks the cylindrical symmetries of the nanotubes and results in further changes in the band structures, including splitting of the subbands $|m\rangle$ in individual nanotubes.[16]

Since, as mentioned above, the perturbation by a transverse electric field is proportional to the nanotube diameter, significantly weaker fields may induce an appreciable change in the band structures of larger-diameter SWNTs. For example, since the SWNTs of 2 to 3 nm in diameter are three to four times larger than the (11,0) CNT, only a third or a quarter of electric field strength (i.e., less than 0.1 V/Å) would be strong enough to make a significant band gap change in the band structure of such large nanotubes. Furthermore, we have found that in small bundles of metallic armchair SWNTs a transverse electric field leads to opening of a band gap similarly as the intertube interaction does.[9] It was previously pointed out that defects in metallic SWNTs can enable large resistance changes induced by a transverse field.[10]

As two atoms move closer, bonding and antibonding states are formed, splitting energy levels. Then the energy level of the bonding state gets lower and the level of the antibonding state gets higher. The same phenomenon occurs in the two-CNT bundle: As two semiconducting CNTs become closer, the top valence bands and bottom conduction bands of two CNTs are split. Due to the energy splitting, one of top valence bands moves up and one of bottom conduction bands moves down. Thus the band gap of the tube bundle decreases. If a transverse external electric field is applied to the bundle, this feature is enhanced and a drastic band gap reduction takes place by the Stark effect. On the other hand, metallic CNTs have degenerate states at the Fermi level. In the armchair CNT, the $\pi$ and $\pi^*$ bands cross each other at the Fermi level. When an applied electric field is introduced, a band mixing of the two bands occurs and the bands are split. Finally, a band gap opens up with a degeneracy breaking.[9] Our results for small bundles of perfect SWNTs show that similar switching behavior can occur because of intertube screening effects and opening of a gap, without the need for defects. This phenomenon thus offers an interesting avenue for applications. High intensity electromagnetic waves will similarly affect the electronic structure of the bundle, adding complex interference effects in the dc and ac conductivities under a high-intensity electric field.

In summary, our first-principles calculations show that small nanotube bundles exhibit attractive electrical characteristics under applied transverse electric field. In particular, the density of states near the Fermi level can undergo significant changes depending on the magnitude and orientation of the electric field with respect to the bundle. The effect is due to the cooperative effect of the SWNT bundle, which enhances the intertube screening induced by the field. As a result, the effective electric field in the intertube region becomes weak, breaking the symmetries of the individual CNTs and altering their electronic structure. Most dramatically, in sufficiently strong fields our results predict the closing of the band gap in bundles consisting of semiconducting CNTs, and the opening of a gap in metallic-CNT bundles. These effects may lead to electrically tunable switching and thus new kinds of CNT-based electronics.

**ACKNOWLEDGMENTS**

This work was supported by Grant No. KHU-20100119 from Kyung Hee University in 2010 (G.K. and Y.-K.K.) and DOE under Grant No. DE-FG02-98ER45685 (J.B.).